\documentclass[11pt,onecolumn,draftcls]{IEEEtran}
\usepackage{amsmath,amssymb,latexsym,cite}
\usepackage{pstricks,graphicx}

\newtheorem{definition}{Definition}
\newtheorem{theorem}{Theorem}

\newtheorem{lemma}{Lemma}

\newcommand{\typ}[1]{\ensuremath{T_{#1}}}

\def\wdep{\mc{W}_{dep}}

\def\prob{\mathbb{P}}
\def\expe{\mathbb{E}}
\def\mbf{\mathbf}
\def\mc{\mathcal}

\newcommand{\norm}[1]{\left\| #1 \right\|}
\newcommand{\kldiv}[2]{\ensuremath{D\left(#1\ \left\|\ #2 \right. \right)}}

\newcommand{\mi}[2]{\ensuremath{I \left( #1 \ \wedge \ #2 \right)}}
\newcommand{\cmi}[3]{\ensuremath{I \left( \left. #1 \ \wedge \ #2 \right| #3 \right)}}
\newcommand{\empMI}[2]{\ensuremath{I\left(#1, #2 \right)}}
\newcommand{\maxvar}[2]{\ensuremath{d_{\max}\left(#1, #2 \right)}}

\newcommand{\usym}{\ensuremath{\mathcal{U}_{\mathrm{sym}}}}
\newcommand{\symchan}{\usym}

\newcommand{\weaklsym}{\ensuremath{\tilde{L}_{\mathrm{sym}}}}
\newcommand{\lsym}{\ensuremath{L_{\mathrm{sym}}}}

\def\capstd{\ensuremath{C_{\mathrm{std}}}}
\def\capdep{\ensuremath{C_{\mathrm{dep}}}}

\def\scostf{l}
\def\scostb{\Lambda}
\def\scostmax{\lambda^{\ast}}
\def\symcost{\lambda}
\def\weaksymcost{\tilde{\lambda}}

\def\lenc{\phi}
\def\ldec{\psi}

\def\stdmaxerr{\varepsilon}

\def\listavgerr{\bar{\varepsilon}_L}

\newcommand{\goodtypes}{\mc{G}_{\eta}(\Lambda)}

\title{State constraints and list decoding for the AVC}
\author{
Anand~D.~Sarwate~\IEEEmembership{Member,~IEEE,} and~Michael~Gastpar~\IEEEmembership{Member,~IEEE}
\thanks{Manuscript received September 2009;
revised XXXXXXXXXXXXXX.  }%
\thanks{A.D. Sarwate is with the Information Theory and Applications Center at the University of California, San Diego, La Jolla CA 92093-0447.  M. Gastpar is with the Department of Electrical Engineering and Computer Sciences, University of California, Berkeley, Berkeley CA 94720-1770 USA.}
\thanks{The work of A.D. Sarwate and M. Gastpar was supported in part by the National Science Foundation under award CCF-0347298.}
}
\date{\today}

\begin{document}

\maketitle

\begin{abstract}
List decoding for arbitrarily varying channels (AVCs) under state constraints is investigated.  It is shown that rates within $\epsilon$ of the randomized coding capacity of AVCs with input-dependent state can be achieved under maximal error with list decoding using lists of size $O(1/\epsilon)$.  Under average error an achievable rate region and converse bound are given for lists of size $L$.  These bounds are based on two different notions of symmetrizability and do not coincide in general.  An example is given that shows that for list size $L$ the capacity may be positive but strictly smaller than the randomized coding capacity.  This behavior is different than the situation without state constraints.
\end{abstract}

\section{Introduction}

The arbitrarily varying channel (AVC) is a model for communication subject to time-varying interference \cite{BlackwellBT:60random}.  The time variation is captured by a channel state parameter and coding schemes for these channels are required to give a guarantee on the probability of error for all channel state sequences.  The AVC is thought of as an adversarial  model in which the channel state is controlled by a \textit{jammer} who wishes to foil the communication between the encoder and decoder.  

This short paper addresses the problem of list-decoding in an AVC when the state sequence is constrained.  The constraint comes by imposing a per-letter cost $l(\cdot)$ on the state sequence and requiring the cost of the state sequence chosen by the jammer for $n$ channel uses to be less than a total budget $\scostb n$.  The randomized and deterministic coding capacity for this AVC variant was found by Csisz\'{a}r and Narayan \cite{CsiszarN:88constraints,CsiszarN:88positivity}.  In particular, they showed that the deterministic coding capacity under average error $\bar{C}_d(\scostb)$ may be positive but strictly smaller than the randomized coding capacity $C_r(\scostb)$.  This is a qualitatively different situation from AVCs without constraints \cite{Ahlswede:78elimination}, where $\bar{C}_d$ is either $0$ or equal to $C_r$.  They also showed that \textit{symmetrizability} as defined by Ericson \cite{Ericson:85exponent} is sufficient for $\bar{C}_d(\scostb)$ to be positive \cite{CsiszarN:88positivity}.

In list-decoding, the decoder is allowed to output a list of $L$ messages and an error is declared only if the list does not contain the transmitted message.  For AVCs without constraints, list-decoding capacities have been investigated under both maximal and average error.  For maximal error, Ahlswede \cite{Ahlswede:73list,Ahlswede:93list} found a quantity $\capdep$ such a rate $\capdep - \epsilon$ is achievable with lists of size $O(1/\epsilon)$.  We extend this result to the situation with cost constraints and define a quantity $\capdep(\scostb)$ such that a rate $\capdep(\scostb) - \epsilon$ is achievable under list-decoding with list size $O(1/\epsilon)$.  This result on maximal error can be used to find the randomized coding capacity of AVCs where the state can depend on the transmitted codeword as well as rateless code constructions \cite{Sarwate:08thesis}.

The average error list-$L$ capacity $\bar{C}_L$ without constraints was found independently by Blinovsky, Narayan, and Pinsker \cite{BlinovskyP:listest,BlinovskyNP:listavc} and Hughes \cite{Hughes:97list}.  These authors defined the symmetrizability $\hat{L}_{\mathrm{sym}}$ of an AVC and showed that there is a constant list size $\hat{L}_{\mathrm{sym}}$ so that for $L \le \hat{L}_{\mathrm{sym}}$ the list-$L$ capacity is $0$ and for $L > \hat{L}_{\mathrm{sym}}$ the list-$L$ capacity is equal to the randomized coding capacity $C_r$.  We show that under state constraints the behavior is qualitatively different.  The ability of the jammer to symmetrize the channel depends on the input distribution $P$ and the cost constraint $\scostb$.  We define two kinds of symmetrizability for list-decoding under state constraints.  We show that for list size $L$ the coding strategy of Hughes \cite{Hughes:97list} can be used with input distributions $P$ such that $L$ is larger than the \textit{weak symmetrizability} $\weaklsym(P,\scostb)$.  We also prove a new converse for input distributions $P$ such that $L$ is smaller than the \textit{strong symmetrizability} $\lsym(P,\scostb)$.

In general, $\lsym(P,\scostb) < \weaklsym(P,\scostb)$, which gives a gap between our achievable region and converse.  Closing this gap seems non-trivial; we conjecture that the converse can be tightened.  However, our results do imply a significant difference between the constrained and unconstrained setting.  Without constraints, the list-$L$ capacity $\bar{C}_L$ is either $0$ or equal to the randomized coding capacity $C_r$.  We show via a simple example that under cost constraints (analogous to \cite{CsiszarN:88positivity}) the list-$L$ capacity $\bar{C}_L(\scostb)$ may be positive but strictly smaller than the randomized coding capacity $C_r(\scostb)$.

\section{Definitions and main results}

We will use calligraphic type for sets and $[M] = \{1, 2, \ldots, M\}$ for integers $M$.  For sets $\mc{X}$ and $\mc{Y}$, the set $\mc{P}(\mc{X})$ is the set of probability distributions on $\mc{X}$, $\mc{P}_n(\mc{X})$ is the set of all distributions of composition $n$, and $\mc{P}(\mc{Y}|\mc{X})$ is the set of all conditional distributions on $\mc{Y}$ conditioned on $\mc{X}$.  For random variables $(X,Y)$ with joint distribution $P_{XY}$ we will write $P_X$ and $P_Y$ for the marginal distributions and $P_{X|Y}$ for the conditional distribution of $X$ given $Y$.  For a distribution $\bar{P} \in \mc{P}(\mc{X}^m)$ we will denote by $P_i$ the $i$-th marginal of $\bar{P}$.  Let $\maxvar{P}{Q}$ be the maximum deviation ($\ell_{\infty}$ distance) between two probability distributions $P$ and $Q$.

\subsection{Channel model and codes}

An AVC is a collection of $\mc{W} = \{W(\cdot | \cdot, s) : s \in \mc{S}\}$ of channels from an input alphabet $\mc{X}$ to an output alphabet $\mc{Y}$ parameterized by a state $s \in \mc{S}$, where all alphabets are finite.  If $\mbf{x} = (x_1, x_2, \ldots, x_n)$, $\mbf{y} = (y_1, y_2, \ldots, y_n)$ and $\mbf{s} = (s_1, s_2, \ldots, s_n)$ are length $n$ vectors, the probability of $\mbf{y}$ given $\mbf{x}$ and $\mbf{s}$ is given by:
	\begin{align}
	W(\mbf{y}| \mbf{x}, \mbf{s}) = \prod_{i=1}^{n} W(y_i | x_i, s_i)~.
	\end{align}
We are interested in the case where there is a bounded cost function $l : \mc{S} \to \mathbb{R}^{+}$ on the jammer.  The cost of an $n$-tuple is
\begin{align}
l(\mbf{s}) = \sum_{k=1}^{n} l(s_k)~.
\end{align}
The state obeys a state constraint $\scostb$ if 
\begin{align}
l(\mbf{s}) \le n \scostb \qquad a.s.~.
\end{align}

An $(n,N,L)$ \textit{deterministic list code} $C$ for the AVC is a pair of maps $(\psi, \phi)$ where the encoding function is $\psi : \{1, 2, \ldots, N\} \to \mc{X}^n$ and the decoding function is $\phi :
\mc{Y}^n \to \{1, 2, \ldots, N\}^L$.  The \textit{rate} of the code is
$R = \log (N/L)$.  The \textit{codebook} is the set of vectors $\{\mathbf{x}_i: 1 \le i \le N\}$, where $\mbf{x}_i = \psi(i)$.  The decoding region for message $i$ is $D_i = \{\mbf{y} : i \in \phi(\mbf{y})\}$.  We will often specify a code by the pairs $\{(\mbf{x}_i, D_i) : i = 1, 2, \ldots, N\}$, with the encoder and decoder implicitly defined.

The \textit{maximal} and \textit{average} error probabilities $\varepsilon_L$ and $\bar{\varepsilon}_L$ are given by
	\begin{align}
	\varepsilon_L 
	&= 
			\max_{\mbf{s} \in \mc{S}^n(\scostb)} 
				\max_{i} \left( 1 - W(D_i | X^n = \mbf{x}_i, \mbf{s}) \right)
			\label{eq:listmaxerr} \\
	\bar{\varepsilon}_L
	&=
	\max_{\mbf{s} \in \mc{S}^n(\scostb)} 
		\frac{1}{N} \sum_{i=1}^{N} \left(1 -  W(D_i | \mbf{x}_i, \mbf{s}) \right)~.
	\label{eq:listavgerr}
	\end{align}
A rate $R$ is called achievable under maximal (average) list-decoding with list size $L$ if for any $\epsilon > 0$ there exists a sequence of $(n,N,L)$ list codes rate at least $R - \epsilon$ whose maximal (average) error converges to $0$.  The list-$L$ capacity is the supremum of achievable rates.  We denote the list-$L$ capacities under maximal and average error by $C_L(\scostb)$ and $\bar{C}_L(\scostb)$, respectively.

\subsection{Symmetrizability and information quantities}

We call a channel $V(y | x_1, x_2, \ldots, x_m)$ from $\mc{X}^m$ to $\mc{Y}$ \textit{symmetric} if for any permutation $\pi$ on $[m]$, 
	\begin{align}
	V(y | x_1, x_2, \ldots, x_m) = V(y | x_{\pi(1)}, x_{\pi(2)}, \ldots, x_{\pi(m)} ) \ \ \ \forall (x_1, x_2, \ldots, x_m, y)~.
	\label{eq:symchandef}
	\end{align}
A channel $U( s | x_1, x_2, \ldots, x_m)$ \textit{symmetrizes} an AVC $\mc{W}$ if
	\begin{align}
	V(y | x, x_1, \ldots, x_m) = \sum_{s \in \mc{S}} W(y | x, s) U( s | x_1, x_2, \ldots, x_m)
	\end{align}
is a symmetric channel.  We denote by $\usym(m)$ the set of channels which symmetrize $\mc{W}$:
	\begin{align}
	\symchan(m) &=  \left\{ U(s | x^m) : 
		V(y | x, x_1, \ldots, x_m)  %
		\textrm{\ is\ symmetric} 
		\right\}~.
		\label{eq:listsymdef}
	\end{align}
Note that $\usym$ is a convex subset of channels $U(s | x_1, \ldots, x_m)$ defined by equality constraints from (\ref{eq:symchandef}).

For a distribution $P \in \mc{P}(\mc{X})$ we define the \textit{strong symmetrizing cost} $\symcost_m(P)$ to be the smallest expected cost of a channel $U(s | x^m)$ that symmetrizes the AVC $\mc{W}$ whose input $\bar{P}(x^m)$ may be correlated but has marginals equal to $P$:
	\begin{align}
	\symcost_m(P) = \min_{U \in \symchan(m)} 
		\max_{\bar{P} \in \mc{P}(\mc{X}^m) : P_i = P} 
		\sum_{x^m} \sum_{s} \bar{P}(x^m) U(s | x^m) l(s)~.
	\label{eq:listsymcost}
	\end{align}
We call an AVC \textit{strongly $m$-symmetrizable} under the constraint $\scostb$ if $\symcost_m(P) \le \scostb$.  We define the \textit{strong symmetrizability} $\lsym(P,\scostb)$ of the channel under input $P$ to be the largest integer $m$ such that $\symcost_{m}(P) < \scostb$.  That is,
	\begin{align}
	\lsym(P,\scostb) = \max \left\{ m : \symcost_{m}(P) < \scostb \right\}~.
	\label{eq:strongsymdef}
	\end{align}
We define the \textit{weak symmetrizing cost} $\weaksymcost_m(P)$ to be the smallest expected cost of a channel $U(s | x^m)$ that symmetrizes the AVC $\mc{W}$ with independent inputs:
		\begin{align}
	\weaksymcost_m(P) = \min_{U \in \symchan(m)}  
		\sum_{x^m} \sum_{s} P^m(x^m) U(s | x^m) l(s)~,
	\label{eq:weaklistsymcost}
	\end{align}
where $P^m$ is the product distribution $P \times P \times \cdots \times P$.  We call an AVC \textit{weakly $m$-symmetrizable} if $\weaksymcost_m(P) \le \scostb$.  Similarly, the \textit{weak symmetrizability} $\weaklsym(P,\scostb)$ is the largest integer $m$ such that $\weaksymcost_{m}(P) < \scostb$.  That is,
	\begin{align}
	\weaklsym(P,\scostb) = \max \left\{ m : \weaksymcost_{m}(P) < \scostb \right\}~.
	\label{eq:weaksymdef}
	\end{align}

For a fixed input distribution $P(x)$ on $\mc{X}$ and channel $V(y | x)$, we will use the notation $\empMI{P}{V}$ to denote the mutual information between the input and output of the channel:
	\begin{align}
	\empMI{P}{V} = \sum_{x,y} V(y | x) P(x) \log \frac{V(y | x) P(x)}{ P(x) \sum_{x'} V(y | x') P(x')}~.
	\label{not:midef}
	\end{align}
	
We define the following two information sets:
	\begin{align}
	\mc{Q}(\scostb) &= 
			\left\{ Q \in \mc{P}(\mc{S}) : 
				\sum_{s} l(s) Q(s) \le \scostb \right\} 
			\label{eq:def:allowQ} \\
		\mc{U}(P, \scostb) &= 
			\left\{ U \in \mc{P}(\mc{S}|\mc{X}) : 
				\sum_{s,x} U(s|x) P(x) l(s) \le \scostb \right\}~.
			\label{eq:def:allowU}
	\end{align}
These in turn can be used to define two information quantities:
	\begin{align}
	\capstd(\scostb) &= \max_{P \in \mc{P}(\mc{X})} \min_{ Q \in \mc{Q}(\scostb) } 
				\empMI{P}{ \sum_{s} W(y | x, s) Q(s) } \\
	\capdep(\scostb) &= \max_{P \in \mc{P}(\mc{X})} \min_{ U \in \mc{U}(P,\scostb) } 
				\empMI{P}{ \sum_{s} W(y | x, s) U(s | x) }~.
	\end{align}

\subsection{Main results}

Our first result extends the strategy of Ahlswede to the case of constrained AVCs under maximal error.

\begin{theorem}[List decoding for maximal error]
\label{thm:maxlist}
Let $\mc{W}$ be an arbitrarily varying channel with state cost function $\scostf(s)$ and cost constraint $\scostb$.  Then for any $\epsilon > 0$ the rate
	\begin{align}
	R = \capdep(\scostb) - \epsilon
	\end{align}
is achievable under maximal error using list decoding with list size 
	\begin{align}
	L = O\left( \frac{1}{\epsilon} \right)~.
	\end{align}
Furthermore, the capacity $C_L(\scostb)$ under maximal error using list decoding with list size $L$ is bounded: 
	\begin{align}
	\capdep(\scostb) - O(L^{-1}) \le C_L(\scostb) \le \capdep(\scostb)~.
	\end{align}
\end{theorem}

The proof is given in Appendix \ref{sec:max}.  This result can be used together with a message authentication strategy \cite{Langberg:04focs} to show that $\capdep(\scostb)$ is the randomized coding capacity of AVCs with input-dependent state \cite{Sarwate:08thesis}.

For average error we can show an achievable rate region and converse bound which in general do not coincide.  Proofs of Theorems \ref{thm:avg:converse} and \ref{thm:avg:achieve} are given in Appendix \ref{sec:avg}.  In both cases the results constrain the set of input distributions in $\mc{P}(\mc{X})$.  The intuition for the converse is that for any codebook with codewords of type $P$, the jammer can choose a symmetrizing channel $U \in \usym(L)$ such that the expected cost under any joint distribution with marginals equal to $P$ is within the cost constraint.  Operationally, the jammer chooses $L$ codewords from the codebook and uses them as inputs to $U$ to generate a state sequence $\mbf{s}$ which satisfies the cost constraints.

\begin{theorem}[Converse for average error]
\label{thm:avg:converse}
Let $\mc{W}$ be an arbitrarily varying channel with state cost function $\scostf(\cdot)$ and cost constraint $\scostb$.  Then we have the following upper bound on $\bar{C}_L(\scostb)$:
	\begin{align}
	\bar{C}_L(\scostb) 
		&\le \max_{P \in \mc{P}(\mc{X}) : \lsym(P,\scostb) < L} \ 
			\min_{ Q \in \mc{Q}(\scostb) } 
			\empMI{P}{ \sum_{s} W(y | x, s) Q(s) }~.
	\end{align}
\end{theorem}

For achievability we extend the coding strategy of Hughes \cite{Hughes:97list} in a manner analogous to \cite{CsiszarN:88positivity} to show an achievable rate for input distributions $P$ such that $L > \weaklsym(P,\scostb)$.

\begin{theorem}[Achievability for average error]
\label{thm:avg:achieve}
Let $\mc{W}$ be an arbitrarily varying channel with state cost function $\scostf(\cdot)$ and cost constraint $\scostb$.  Then we have the following lower bound on $\bar{C}_L(\scostb)$:
	\begin{align}
	\bar{C}_L(\scostb) 
	    &\ge \max_{P \in \mc{P}(\mc{X}) : \weaklsym(P,\scostb) < L} \ 
	    		\min_{ Q \in \mc{Q}(\scostb) } 
			\empMI{P}{ \sum_{s} W(y | x, s) Q(s) }~.
	\end{align}
If $P^{\ast}$ is the maximizing input distribution for $\capstd(\scostb)$, then for list size $L > \weaklsym(P^{\ast},\scostb)$ we have
	\begin{align}
	\bar{C}_L(\scostb) = \capstd(\scostb)~.
	\end{align}
\end{theorem}

\section{Example and discussion}

We will now show via an example that the behavior of list-decoding under average error with state constraints is qualitatively different from that without constraints.  In particular when the jammer must satisfy a constraint $\scostb < \infty$, positive rates may be achievable with list sizes that are smaller than the unconstrained symmetrizability, and for a fixed list size the list-$L$ capacity may be positive but strictly smaller than the randomized coding capacity.  Let the input $\mc{X} = \{0,1\}$, state $\mc{S} = \{0,1,\ldots, \sigma\}$ and the channel be defined by:
	\begin{align}
	Y = X + S~.  \label{eq:exampleChan}
	\end{align}
We will consider a quadratic cost function $l(s) = s^2$.

Without constraints, Hughes \cite{Hughes:97list} has found that the randomized capacity is
	\begin{align}
	C_r(\infty) = - \log \cos \frac{\pi}{\sigma + 3}~.
	\label{eq:ex:uncons}
	\end{align}
He also showed that for unconstrained AVCs the list-$L$ capacity obeys a strict threshold : 
	\begin{align}
	C_L(\infty) = \left\{ \begin{array}{ll}
		- \log \cos \frac{\pi}{\sigma + 3} & L > \sigma \\
		0 								 & L \le \sigma
		\end{array}
		\right.
	\label{eq:list:uncons}
	\end{align}
We are interested in the case when there is a cost constraint $\scostb$ on the jammer.  We must calculate the minimum mutual information for different input distributions:
	\begin{align}
	\empMI{P}{\scostb} = \min_{Q \in \mc{P}(\mc{S}) : \expe_Q[l(s)] \le \scostb}
		\mi{X}{Y}~.
	\end{align}
The randomized-coding capacity under the cost constraint $\scostb$ is the max of $\empMI{P}{\scostb}$ over $P$.
	\begin{align}
	C_r(\scostb) = \max_{P \in \mc{P}(\mc{X})} \empMI{P}{\scostb}~.
	\label{eq:ex:cr}
	\end{align}
These calculations can be easily done numerically.  

To calculate the symmetrizability constraints, note that the because the channel (\ref{eq:exampleChan}) is deterministic, the symmetry constraints imply that any channel $U\in \usym$ must also be symmetric.  Therefore $U(s | x_1, x_2, \ldots, x_L)$ is only a function of the type of $(x_1, x_2, \ldots, x_L)$.  Let $t$ denote this type.  We now view $\usym$ as containing channels $U(s | t)$.  Note that for $y = 0$ we have
	\begin{align}
	\sum_{s} W(0 | 0, s) U(s | t) = U(0 | t)~,
	\end{align}
and by the symmetry constraint we have
	\begin{align}
	U(0 | t) = 0 \qquad t = 1, 2, \ldots, L~.
	\label{eq:symCon1}
	\end{align}
Similarly, for $y = \sigma + 1$ we have
	\begin{align}
	U(\sigma | t) = 0 \qquad	 t = 0, 1, \ldots, L-1~.
	\label{eq:symCon2}
	\end{align}
Finally, for $y = 1, 2, \ldots, \sigma$ we have
	\begin{align}
	\sum_{s} W(y | 0, s) U(s | t) &= U(y | t) \\
	&= \sum_{s} W(y | 1, s) U(s | t-1) \\
	&= U(y-1 | t-1) \qquad y = 1, 2, \ldots, \sigma, \ \ t = 1,2,\ldots, L
	\label{eq:symCon3}
	\end{align}
The conditions (\ref{eq:symCon1}), (\ref{eq:symCon2}), and (\ref{eq:symCon3}) characterize the linear symmetry constraints in $\usym$.

Thus for each input distribution $P$ we can find
	\begin{align}
	f(P) = \min_{U \in \usym} \sum_{s, t} l(s) U(s | t) \binom{L}{t} P(0)^{L-t} P(1)^t~.
	\end{align}
This is a simple linear program.  To calculate the strong $L$-symmetrizing cost, note that the set of all joint distributions $\bar{P}(x_1^L)$ with marginals equal to $P$ is
also a convex set defined by linear equality constraints.  If we let 
	\begin{align}
	\tau(\bar{P},t) = \sum_{x_1^L : T_{\mbf{x}} = t/L} \bar{P}(x_1^L)~,
	\end{align}
be the probability of a type-$t$ sequence under $\bar{P}$, it is simple to numerically evaluate
	\begin{align}
	g(P) = \max_{\bar{P}} \min_{U \in \usym} \sum_{s,t} l(s) U(s | t) \tau(\bar{P},t)~.
	\end{align}

\begin{figure}[htb]
\centering 
\includegraphics[width=3.5in]{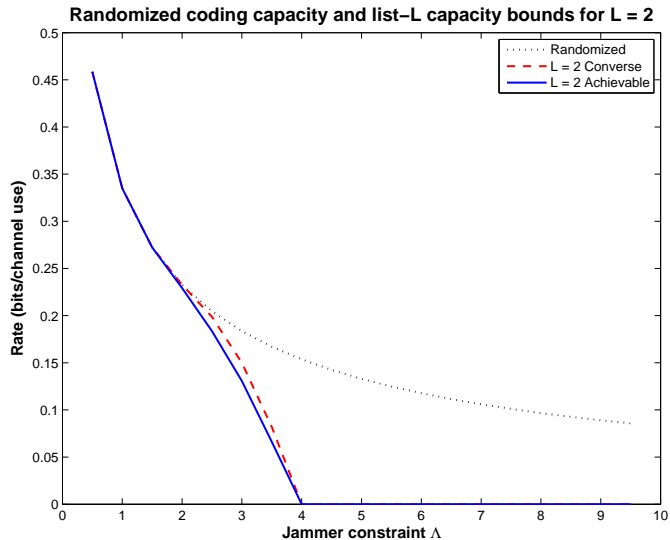} 
\caption{Randomized coding capacity $C_r(\scostb)$ and bounds on list-$L$ capacity $\bar{C}_L(\scostb)$ versus the state constraint $\scostb$ for $L = 2$.} 
\label{fig:ListRates2} 
\end{figure} 

\begin{figure}[htb]
\centering 
\includegraphics[width=3.5in]{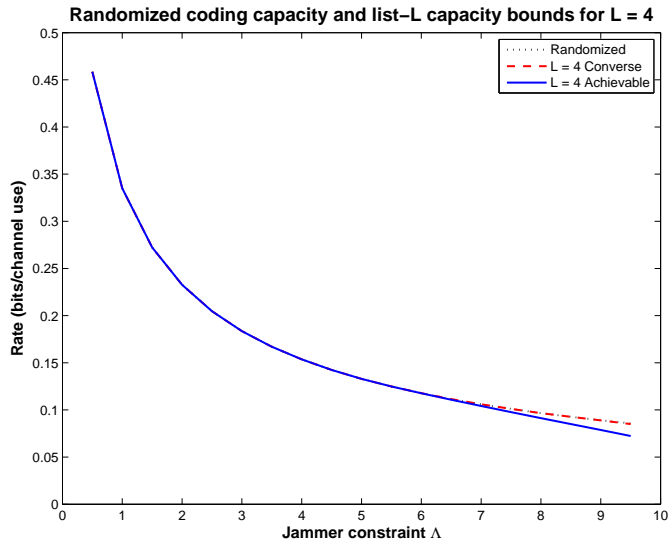} 
\caption{Randomized coding capacity $C_r(\scostb)$ and bounds on list-$L$ capacity $\bar{C}_L(\scostb)$ versus the state constraint $\scostb$ for $L = 4$.} 
\label{fig:ListRates4} 
\end{figure} 

We calculated the achievable rates and converse bounds for $\sigma = 8$, and the results are shown for list sizes $L = 2$ and $L = 4$ in Figures \ref{fig:ListRates2} and \ref{fig:ListRates4}.  For state constraint $\scostb$, the randomized coding capacity $C_r(\scostb)$ in (\ref{eq:ex:cr}) is given by the dotted line.  The achievable rate of Theorem \ref{thm:avg:achieve} is shown by the solid line, and the converse bound of Theorem \ref{thm:avg:converse} by the dashed line.  These two curves are given by restricting the optimization over $P$ in the right side of (\ref{eq:ex:cr}).

When $\scostb = \infty$, the randomized coding capacity of this channel is given by (\ref{eq:ex:uncons}) and is $0.0597$ bits/channel use.  Therefore, when $\scostb = \infty$, the result in (\ref{eq:list:uncons}) shows that the the list-$L$ capacity is $0$ for $L < 8$ and equal to $0.0597$ for $L > 8$. That is, when the jammer is unconstrained, no positive rate is achievable under average error using list decoding with list size smaller than $8$.  However, from Figures \ref{fig:ListRates2} and \ref{fig:ListRates4} we can see that when $\scostb < \infty$ we can achieve positive rates for list sizes $L$ smaller than 8.  However, for a range of $\scostb$, the randomized coding capacity is achievable using lists of size 2 or 4.  Figure \ref{fig:ListRates2} also illustrates another fundamental difference between list-decoding with state constraints and list-decoding without constraints: for a range around $\scostb = 3$, the list-$2$ capacity $\bar{C}_2(\scostb)$ is positive but strictly smaller than the randomized coding capacity $C_r(\scostb)$.

In general, we conjecture that the converse region of Theorem \ref{thm:avg:converse} is not tight and that a stronger converse could be shown.  The strong symmetrizing cost in (\ref{eq:listsymcost}) allows optimization over all joint distributions with the same marginals.  The converse proof uses a jamming strategy corresponding to taking a random set of $L$ codewords from the codebook as inputs to a symmetrizing channel $U(s | x^L)$ to generate the state sequence.  The strong symmetrizing cost is a conservative bound on the cost of such a strategy.  It may be that techniques such as \cite{ShamaiV:97good} could improve this bound; we leave this for future work.  Our results here establish that the behavior of list-decoding for constrained AVCs is fundamentally different than the unconstrained case, much like the situation for list size $1$.

\appendices

\section{Maximal Error \label{sec:max}}

Using now-standard typicality arguments we can show the existence of list-decodable codes for maximal error with exponential list size.  The codebook is the entire set of typical sequences $\typ{P}$ and the list is the union of $\epsilon$-shells under the different state sequences.  The decoder outputs a list that is the union of shells.  Let
	\begin{align}
		\wdep(P, \scostb) &= 
		\left\{V(y|x) : V(y|x) = 
			\sum_{s} W(y|x,s) U(s|x), 
				\ \ U(s|x) \in \mc{U}(P, \scostb) \right\}~.
		\label{eq:dephull}
	\end{align}

\begin{proof}[Proof of Theorem \ref{thm:maxlist}]
The converse argument follows by choosing $\mbf{s}$ according to the minimizing distribution $U(s|x)$ in $\mc{U}(P,\scostb)$.  To show the achievable rate, without loss of generality, suppose that the distribution $P$ maximizing $\capdep(\scostb)$ is in $\mc{P}_n(\mc{X})$ and consider the set $\typ{P}$ of all sequences of length $n$ of type $P$ (if not we can always approach the optimal $P$ with large $n$).  For any $V(y|x)$ we define $V'(x | y)$ from $V(y|x)P(x)$ via the Bayes rule.  The $(V',\epsilon)$-shell of typical $\mbf{x}$ sequences around a $\mbf{y}$ is:
	\begin{align}
	T_{V'}^{\epsilon}(\mbf{y}) 
		= \left\{ \mbf{x} \in \typ{P} 
			: \maxvar{  \typ{\mbf{x}{y}}  }{  V' \typ{\mbf{y}} } < \epsilon
			\right\}~. 
	\end{align}
Then
	\begin{align}
	\frac{1}{n} \log \left| T_{V'}^{\epsilon}(\mbf{y}) \right|
		\le H_{V'\typ{\mbf{y}}}(X | Y) + O( \epsilon \log \epsilon^{-1} )~,
	\end{align}
where the subscript on $H$ indicates the the joint distribution under which to take the mutual information.

Now, for a fixed $\mbf{x} \in \typ{P}$ and $\mbf{s}$ with $l(\mbf{s}) \le n \scostb$, we define an empirical forward channel
	\begin{align}
	V_{\mbf{xs}}(y | x) = \sum_{s} W(y | x, s) \frac{ N(x,s | \mbf{x},\mbf{s}) }{ N(x | \mbf{x})}~.
	\end{align}
Note that $V_{\mbf{xs}} \in \wdep(P,\scostb)$.  For a fixed received codeword $\mbf{y}$, define the set of channels consistent with $\mbf{y}$ as:
	\begin{align}
	\mc{V}_P^{\delta}(\mbf{y}) 
		= \left\{  V \in \wdep(P,\scostb) \cap \mc{P}_n(\mc{Y}|\mc{X}) 
			: \maxvar{ \sum_{y} V(y|x) P(x) }{ \typ{\mbf{y}} } < \delta	
			\right\}~.
	\end{align}

Consider the set
	\begin{align}
	\mc{A}(\mbf{y}) = \bigcup_{ V \in \mc{V}_P^{\delta}(\mbf{y}) }
		T_{V'}^{( |\mc{X}| + 1| ) \delta}(\mbf{y})~.
	\end{align}
Standard typicality arguments show that if $\mbf{x}$ generated $\mbf{y}$ via some $\mbf{s}$ satisfying the cost constraint, then with probability $1 - \exp(-n E(\delta))$, we have $\mbf{x} \in \mc{A}(\mbf{y})$.  Furthermore:
	\begin{align}
	\frac{1}{n} \log |\mc{A}(\mbf{y})| 
		\le 
		\min_{V \in \wdep(P,\scostb)} H_{V(y|x)P(x)}(X | Y)
		+ O( \delta \log \delta^{-1} )~.
	\label{eq:biglistsize}
	\end{align}

Note that we can view an encoding into all of $\typ{P}$ and decoding into $\mc{A}(\mbf{y})$ as a list-decodable code with $2^{n H(P)}$ codewords and list size (\ref{eq:biglistsize}).  To arrive at the desired code we can sample a set $\mc{B} = \{\mbf{x}(i)\}$ of $2^{n (\capdep(\scostb) - \epsilon)}$ codewords from this $\typ{P}$ uniformly at random and say the decoder outputs $\mc{A}(\mbf{y}) \cap \mc{B}$.  We must show this set has at most $L = O(1/\epsilon)$ codewords with high probability.  

Let $R = \capdep(\scostb) - \epsilon$.  For each $\mbf{y}$, the probability that any codeword of $B$ is in $\mc{A}(\mbf{y})$ is upper bounded by $|\mc{A}(\mbf{y})|/|\typ{P}|$, so from (\ref{eq:biglistsize}) we see
	\begin{align}
	\prob\left( \mbf{x}(i) \in \mc{A}(\mbf{y}) \right) 
		\le \exp\left(
			- n \left(  \capdep(\scostb) - O(\delta \log \delta^{-1}) \right) 
			\right)~.
	\end{align}
Since codewords are selected independently, we can bound the chance that a fraction $L \cdot 2^{-nR}$ of the $2^{nR}$ codewords end up in $\mc{A}(\mbf{y})$ using Sanov's theorem \cite[Theorem 12.4.1]{CoverThomas} 
	\begin{align}
	\prob\left( |\mc{A}(\mbf{y}) \cap \mc{B}| > L \right)
	\le
	\exp\left( 
		- 2^{nR} \kldiv{ 
				L 2^{- nR} 
			}{ 
				2^{ -n (\capdep(\scostb) - O(\delta \log \delta^{-1})} 
			}
		+ h\log (2^{n R} + 1) \right)
	\label{eq:sanovbound}
	\end{align}
Now we can bound the term $2^{nR} \kldiv{\cdot}{\cdot}$:
	\begin{align}
	-L \log \frac{L}{2^{n (\epsilon - O(\delta \log \delta^{-1})) }}
	- 2^{nR} (1 - L 2^{-nR}) 
		\log \frac{ 
				1 - L 2^{-nR} 
			}{ 
				1 -  2^{-n (R + \epsilon - O(\delta \log \delta^{-1}) ) } 
			}
	 \\
	& \hspace{-2in} 
	\le
		- n L \left( \epsilon - O(\delta \log \delta^{-1}) \right)
		- L \log L
		+ 2 L~.
	\label{eq:sanovreduce}
	\end{align}
We can pick $\delta$ such that $O(\delta \log \delta^{-1}) < \epsilon/2$ by choosing $n$ sufficiently large.  Then substituting (\ref{eq:sanovreduce}) in (\ref{eq:sanovbound}), upper bounding $R < \log |\mc{Y}|$, and taking a union bound over all $\mbf{y}$ we have: 
	\begin{align}
	\prob\left( \exists \mbf{y} \ : \ |\mc{A}(\mbf{y}) \cap \mc{B}| > L \right)
		\le \exp\left( - n \left(  
			L \epsilon/2 + 2 \log |\mc{Y}| \right) - L \log L + 2 L \right)~.
	\end{align}
For sufficiently large $n$ choosing $L > \lceil \frac{4 \log |\mc{Y}|}{\epsilon} \rceil$ makes the exponent negative, showing that with high probability the random selection will produce an $(n, 2^{nR}, L)$ list-decodable code under maximal error whose error is bounded by $1 - \exp(-n E(\delta))$.
\end{proof}

\section{Average Error \label{sec:avg}}

\subsection{Facts about symmetrizability}

The following theorem shows that if $I(P)$ is positive, then $\weaklsym(P,\scostb)$ is finite.  In particular, since $\empMI{P^{\ast}}{\scostb}$ is finite, the theorem implies that if $\capstd(\scostb) > 0$, then $\weaklsym(P^{\ast},\scostb) < \infty$.  The proof follows straightforwardly from the results of \cite{Hughes:97list}.

\begin{lemma}[Finite symmetrizability]
\label{lem:avg:finitesym}
Let $\mc{W}$ be an arbitrarily varying channel with state cost function $\scostf(\cdot)$.  If $\capstd(\scostb) = 0$ then $\lsym(P,\scostb) = \infty$ for all $P$.  If $\capstd(\scostb) > 0$ then 
	\begin{align}
	\weaklsym(P,\scostb) \le \frac{\log(\min( |\mc{Y}|, |\mc{S}| ))}{\empMI{P}{\scostb}}
	\end{align}
for all $P$ such that $\empMI{P}{\scostb} > 0$.
\end{lemma}

\subsection{Achievability under average error \label{sec:avg:ach}}

Given a $P$ that is not weakly $L$-symmetrizable, we can use the coding scheme of Hughes \cite{Hughes:97list} modified in the natural way suggested by Csisz\'{a}r and Narayan \cite{CsiszarN:88positivity} for list size $1$.  The codebook consists of $N$ constant-composition codewords drawn uniformly from the codewords of type $P$.  In order to describe the decoding rule we will use, we define the set 
\begin{align}
\goodtypes = \{P_{XSY} \in \mc{P}(\mc{X} \times\mc{S} \times \mc{Y}) : \kldiv{ P_{XSY} }{ P_X \times P_S \times W } \le \eta,\ \expe[l(s)] \le \scostb \}~,
\label{eq:avg:gtypedef}
\end{align}
where
	\begin{align}
	(P_X \times P_S \times W)(x,s,y) = P_X(x) P_S(s) W(y | x, s)~.
	\end{align}
The set $\goodtypes$ contains joint distributions which are close to those generated from the AVC $\mc{W}$ via independent inputs with distribution $P_X$ and $P_S$.

\begin{definition}[Decoding rule]
\label{def:avg:decode}
Let $\mbf{x}_1, \mbf{x}_2, \ldots, \mbf{x}_N$ be a given codebook and suppose $\mbf{y}$ was received.  Let $\ldec(\mbf{y})$ denote the list decoded from $\mbf{y}$.  Then put $i \in \ldec(\mbf{y})$ if and only if there exists an $\mbf{s} \in \mc{S}^n(\scostb)$ such that
\begin{enumerate}
\item $T_{\mbf{x}_i \mbf{s} \mbf{y}} \in \goodtypes$, and
\item for every set of $L$ other distinct codewords $\{\mbf{x}_{j} : j \in J,\ J \subset [N]\setminus\{i\},\ |J| = L\}$ such that there exists a set $\{\mbf{s}_{j} : \mbf{s}_{j} \in \mc{S}^n(\scostb),\ j \in J \}$ with $T_{\mbf{x}_{j} \mbf{s}_j \mbf{y}} \in \goodtypes$ for all $j \in J$ we have
\begin{align}
\cmi{YX}{X^L}{S} \le \eta~,
\label{eq:avg:decbound}
\end{align}
where $P_{Y X X^L S}$ is the joint type of $(\mbf{y}, \mbf{x}_i, \{\mbf{x}_j : j \in J\}, \mbf{s})$.
\end{enumerate}
\end{definition}

An interpretation of this rule is that the decoder outputs a list of codewords $\{\mbf{x}_i\}$ each having a ``good explanation'' $\{\mbf{s}_i\}$.  A ``good explanation'' is a state sequence that plausibly could have generated the observed output $\mbf{y}$ (condition 1) and makes all other $L$-tuples of codewords seem independent of the codeword and output (condition 2).  The only thing to prove is that this decoding rule is unambiguous.  The key is to show that no tuple of random variables $(Y, X^{L+1}, S^{L+1})$ can satisfy the conditions of the decoding rule.  This in turn shows that for sufficiently large $n$, no set of $L+1$ \textit{codewords} can satisfy the conditions of the decoding rule.  Therefore, for sufficiently large blocklengths, the decoding rule will only output $M$ or fewer codewords.

\begin{lemma}
\label{lem:avg:sep}
Let $\beta > 0$, $\mc{W}$ be an AVC with state cost function $l(\cdot)$ and constraint $\scostb$, $P \in \mc{P}(\mc{X})$ with $I(P,\scostb) > 0$ and $\min_x P(x) \ge \beta$, and $M = \weaklsym(P,\scostb) + 1$.  For any $\alpha > 0$ and every collection of distributions $\{ U_i \in \mc{P}(\mc{X}^{M} \times \mc{S}) : i = 1, 2, \ldots, M\}$ such that 
	\begin{align}
	\sum_{x^{M+1}, s} P(x_i) U_i(x^{M}_{-\{i\}}, s) l(s) 
		\le \weaksymcost_{M}(P) - \alpha
	\label{eq:avg:constr_sep}
	\end{align}
for all $i = 1, 2, \ldots, M+1$, there exists a $\zeta > 0$ such that 
\begin{align}
\max_{j \ne i} \sum_{y, x^{M+1}} \left| 
	\sum_{s} W(y | x_i, s) U_i(x^{M+1}_{-\{i\}}, s) P(x_i) 
	- \sum_{s} W(y | x_j, s) U_j(x^{M+1}_{-\{j\}}, s) P(x_j) 
	\right| \ge \zeta~.
\label{nonsym_sep}
\end{align}
\end{lemma}

\begin{proof}
Note that the outer sum in (\ref{nonsym_sep}) is over all $x^{M+1}$.
Define the function $V_k : \mc{X}^{M+1} \times \mc{S} \to \mathbb{R}$ by:
	\begin{align}
	V_k(x^{M+1}, s) = U_k(x^{M+1}_{-\{k\}}, s)~.
	\end{align}
Let $\Pi_{M+1}$ be the set of all permutations of $[M+1]$ and for $\pi \in \Pi_{M+1}$ let $\pi_i$ be the image of $i$ under $\pi$.  Then
\begin{align}
&\max_{j \ne i} \sum_{y, x^{M+1}} \left| 
	\sum_{s} W(y | x_i, s) V_i(x^{M+1}, s) P(x_i) 
	- \sum_{s} W(y | x_j, s) V_j(x^{M+1}, s) P(x_j) 
	\right| \nonumber \\
	& \hspace{0.1cm}
= \max_{j \ne i} \sum_{y, x^{M+1}} \Bigg| 
	\sum_{s} W(y | x_i, s) V_{\pi_i}( \pi(x^{M+1}), s) P(x_i) 
	 \nonumber \\
&\hspace{2in} 
	- \sum_{s} W(y | x_j, s) V_{\pi_j}( \pi(x^{M+1}), s) P(x_j) 
	\Bigg|~.
\end{align}
We can lower bound this by averaging over all $\pi \in \Pi_{M+1}$ :
	\begin{align}
	\max_{j \ne i} 
		& 
		\sum_{y, x^{M+1}} \frac{1}{(M+1)!} 
			\sum_{\pi \in \Pi_{M+1}} \left| 
				\sum_{s} W(y | x_i, s) V_{\pi_i}(\pi(x^{M+1}), s) P(x_i) 
	\rule{0pt}{15pt} \right. \nonumber \\
	& \hspace{2.5in}
	\left. \rule{0pt}{15pt}
	- \sum_{s} W(y | x_j, s) V_{\pi_j}(\pi(x^{M+1}), s) P(x_j) 
	\right|~.
	\label{eq:avg_sep:perm}
\end{align}

Define the average
\begin{align*}
\bar{V}(x^{M+1}_{-\{i\}}, s) &= \frac{1}{(M+1)!} \sum_{\pi \in \Pi_{M+1}} V_{\pi_i}(\pi(x^{M+1}), s) \\
&= \frac{1}{(M+1)!} \sum_{l=1}^{M+1} \sum_{\pi \in \Pi_{M+1} : \pi_i = l} U_l( \pi( x^{M+1} )_{-\{\pi_i\}}, s) \\
&= \frac{1}{(M+1)!} \sum_{l=1}^{M+1} \sum_{\sigma \in \Pi_{M}} U_l(\sigma( x^{M+1}_{-\{i\}}), s)~.
\end{align*}
Note that $\bar{V}$ is a symmetric function for all $s$.

Now we use the convexity of $|\cdot|$ to pull the averaging inside
the absolute value to get a further lower bound on (\ref{eq:avg_sep:perm}) by substituting in $\bar{V}$.
	\begin{align}	
	F(\bar{V},P) 
	&= 
	\max_{j \ne i} \sum_{y, x^{M+1}} 
		\left| 
			\sum_{s} W(y | x_i, s) \bar{V}(x^{M+1}_{-\{i\}}, s) P(x_i) 
	\rule{0pt}{15pt} \right. \nonumber \\
	& \hspace{2.5in}
	\left. \rule{0pt}{15pt}
		- \sum_{s} W(y | x_j, s) \bar{V}(x^{M+1}_{-\{j\}}, s) P(x_j) 
	\right|~.
\label{symm_lbound}
\end{align}
The function $F(\bar{V},P)$ is continuous function on the compact set of symmetric distributions $\{\bar{V}\}$ and the set of distributions $P$ with $\min_x P(x) \ge \beta$, so it has a minimum $\zeta = F(\bar{V}^{\ast}, P^{\ast})$ for some $(\bar{V}^{\ast}, P^{\ast})$.  We will prove that $\zeta > 0$ by contradiction.

Suppose $F(\bar{V}^{\ast}, P^{\ast}) = 0$.  Then
	\begin{align*}
	\sum_{s} W(y | x_i, s) \bar{V}^{\ast}(x^{M+1}_{-\{i\}}, s) P^{\ast}(x_i) 
	= \sum_{s} W(y | x_j, s) \bar{V}^{\ast}(x^{M+1}_{-\{j\}}, s) P^{\ast}(x_j)~.
	\end{align*}
So
	\begin{align*}
	\sum_{y} \sum_{s} W(y | x_i, s) \bar{V}^{\ast}(x^{M+1}_{-\{i\}}, s) P^{\ast}(x_i) 
	&= \sum_{y} \sum_{s} W(y | x_j, s) \bar{V}^{\ast}(x^{M+1}_{-\{j\}}, s) P^{\ast}(x_j) \\
\bar{V}^{\ast}(x^{M+1}_{-\{i\}}) P^{\ast}(x_i) &= \bar{V}^{\ast}(x^{M+1}_{-\{j\}}) P^{\ast}(x_j)~,
	\end{align*}
which implies (see \cite[Lemma A3]{Hughes:97list}) that for all $j$:
	\begin{align*}
	\bar{V}^{\ast}(x^{M+1}_{-\{j\}}) P^{\ast}(x_j) = P^{\ast (M+1)}(x^{M+1})~.
	\end{align*}
Therefore
	\begin{align}
	\sum_{s} W(y | x_1, s) \bar{V}^{\ast}(s | x_{2}^{M+1})~.
	\label{eq:avg_sep:sym}
	\end{align}
is symmetric in $(x_1, x_2, \ldots, x_{M+1})$.  Therefore $\bar{V}^{\ast}(s | x_{2}^{M+1}) \in \symchan(M+1)$.  From the definition of $\weaksymcost_{M}(P)$ in (\ref{eq:weaklistsymcost}) we see that
	\begin{align}
	\sum_{x^{M+1}, s} \bar{V}^{\ast}(x^{M}_{-\{i\}}, s) P(x_i) l(s) &\ge \weaksymcost_{M}(P)~.
	\end{align}
But from (\ref{eq:avg:constr_sep}), and the definition of $\bar{V}$ we see that the $\{U_i\}$ must be chosen such that
	\begin{align}
	\sum_{x^{M+1}, s} \bar{V}^{\ast}(x^{M}_{-\{i\}}, s) P(x_i) l(s) &\le \weaksymcost_{M}(P) - \alpha~.
	\label{sym_constr}
	\end{align}
Therefore we have a contradiction and the minimum $\zeta$ of $F(\bar{V},P)$ must be greater than $0$.  Equation (\ref{nonsym_sep}) follows.
\end{proof}

The next lemma shows that for a sufficiently small choice of the threshold $\eta$ in the decoding rule there are no random variables that can force the decoding rule to output a list that is too large.  The proof follows from Lemma \ref{lem:avg:sep} in the same way as in \cite{Hughes:97list}.

\begin{lemma}
\label{lem:avg:nambig}
Let $\beta > 0$, $\mc{W}$ be an AVC with state cost function $l(\cdot)$ and constraint $\scostb$, $P \in \mc{P}(\mc{X})$ with $\min_x P(x) \ge \beta$, and $M = \weaklsym(P,\scostb) + 1$.  Then there exists an $\eta > 0$ sufficiently small such that no tuple of rv's $(Y, X^{M+1}, S^{M+1})$ can simultaneously satisfy
\begin{align}
\min_{x} P(x) &\ge \beta \label{eq:avg:tuple1}\\
P_{X_i} &= P \\
P_{Y X_i S_i} &\in \goodtypes \label{eq:avg:tuple3} \\
\cmi{ Y X_i }{ X^{M+1}_{-\{i\}} }{ S_i } &\le \eta \ \ \ 1 \le i \le M+1
\label{eq:avg:tuple4}
\end{align}
\end{lemma}

\begin{proof}[Proof of Theorem \ref{thm:avg:achieve}]
Given Lemma \ref{lem:avg:nambig} the theorem follows from Lemma 3 of \cite{Hughes:97list}.
\end{proof}

\subsection{Converse \label{sec:avg:conv}}

The key idea in the converse is to show that for a codebook with codewords whose types are symmetrizable and close to a fixed symmetrizable type $P$, then the jammer has a strategy that keeps the error bounded away from $0$.  The rest follows from approximation and covering arguments.

\begin{lemma}[Approximating joint distributions]
\label{lem:dist_approx}
Let $\mc{X}$ be a finite set with $|\mc{X}| \ge 2$.  For any $\epsilon > 0$ and probability distribution $P$ on $\mc{X}$ there exists a $\delta > 0$ such that for any collection of distributions $\{P_i \in \mc{P}(\mc{X}) : i \in [L]\}$ satisfying
	\begin{align}
	\maxvar{ P_i }{ P } < \delta \qquad \forall i
	\label{eq:avg:closemarg}
	\end{align}
and any joint distribution $\bar{P}(x_1, x_2, \ldots, x_L)$ with
	\begin{align}
	\sum_{x_j : j \ne i} \bar{P}(x_1, x_2, \ldots, x_L) = P_i(x_i) \qquad \forall i,\ x_i \in \mc{X} 
	\label{eq:avg:oldmarg}
	\end{align}
there exists a joint distribution $\hat{P}(x_1, x_2, \ldots, x_L)$ such that
	\begin{align}
	\sum_{x_j : j \ne i} \hat{P}(x_1, x_2, \ldots, x_L) = P(x_i) \qquad \forall i,\ x_i \in \mc{X}
	\label{eq:avg:newmarg}
	\end{align}
and
	\begin{align}
	\maxvar{ \bar{P} }{ \hat{P} } < \epsilon~.
	\label{eq:avg:closejoint}
	\end{align}
\end{lemma}

\begin{proof}[Proof of Lemma \ref{lem:dist_approx}]
Fix $\epsilon > 0$ and $P$.  We consider two cases depending on whether $\min_{x \in \mc{X}} P(x) = 0$ or not.  

\textbf{Case 1.}  First suppose $\min_{x \in \mc{X}} P(x) = \beta > 0$.  Consider a set of distributions $\{P_i : i \in [L]\}$ satisfying (\ref{eq:avg:closemarg}) and let $\bar{P}(x_1^L)$ be a joint distribution satisfying (\ref{eq:avg:oldmarg}).  We treat probability distributions as vectors in $\mathbb{R}^{|\mc{X}|^L}$.   We can construct a distribution $\hat{P}$ satisfying (\ref{eq:avg:newmarg}) and (\ref{eq:avg:closejoint}) in two steps:  first we project $\bar{P}$ onto the set of all vectors whose entries sum to $1$ and satisfy (\ref{eq:avg:newmarg}), and then we find a $\hat{P}$ close to this projection which is a proper probability distribution.

Let $\mc{B}$ be the subspace of $\mathbb{R}^{|\mc{X}|^L}$ of all vectors $P'$ satisfying the marginal constraints (\ref{eq:avg:newmarg}) as well as the sum probability constraint
	\begin{align}
	\sum_{x_1^L} P'(x_1^L) = 1~.
	\label{eq:avg:totprob}
	\end{align}
We can summarize these linear constraints in the matrix form
	\begin{align}
	A P' = b'~,
	\end{align}
where $A$ contains the coefficients on the left-hand sides of the constraints (\ref{eq:avg:newmarg}) and (\ref{eq:avg:totprob}) and  $b'$ has the right-hand sides.  We can assume $A$ has full row-rank by removing linearly dependent constraints.  Note that the distribution $\bar{P}$ satisfies
	\begin{align}
	A \bar{P} = \bar{b}~,
	\end{align}
where $\bar{b}$ has the right-hand sides of (\ref{eq:avg:oldmarg}) instead of (\ref{eq:avg:newmarg}).

Now let $\tilde{P}$ be the Euclidean projection of $\bar{P}$ onto the subspace $\mc{B}$ : 
	\begin{align}
	\tilde{P} = \bar{P} + A^T (A A^T)^{-1} (b' - A \bar{P})~.
	\end{align}
The error in the projection is 
	\begin{align}
	\bar{P} - \tilde{P} &= A^T (A A^T)^{-1} (A \bar{P} - b') \\
	&= A^T (A A^T)^{-1} (\bar{b} - b')~.
	\end{align}
From (\ref{eq:avg:closemarg}) we can see that all elements of $(\bar{b} - b')$ are in $(-\delta,\delta)$.  Since the rows of $A$ are linearly independent, the singular values of $A$ are strictly positive and a function of $|\mc{X}|$ and $L$ only.  Therefore there is a function $\mu_1(|\mc{X}|,L)$ such that 
	\begin{align}
	\norm{A^T (A A^T)^{-1} (\bar{b} - b')}_2 < \mu_1(|\mc{X}|,L) \cdot \delta~.
	\end{align}
Since $|\mc{X}|$ is finite there is a function $\mu_2(|\mc{X}|,L)$ such that
	\begin{align}
	\maxvar{\tilde{P}(x_1^L)}{\bar{P}(x_1^L)}
	< \mu_2(|\mc{X}|,L) \cdot \delta~.
	\end{align}
If the resulting $\tilde{P}$ from this first projection has all nonnegative entries, then we set $\hat{P} = \tilde{P}$ and choose $\delta$ sufficiently small so that $\mu_2(|\mc{X}|,L) \cdot \delta < \epsilon$.

If $\tilde{P}$ has entries that are not in $[0,1]$ then it is not a valid probability distribution.  However, since $\bar{P}$ is a probability distribution, we know that
	\begin{align}
	\min_{x_1^L} \tilde{P}(x_1^L) > -\mu_2(|\mc{X}|,L) \cdot \delta~.
	\end{align}
Let $P^L$ be the joint distribution on $\mc{X}^L$ with independent marginals $P$:
	\begin{align}
	P^L(x_1, \ldots, x_L) = P(x_1) \cdots P(x_L)~.
	\end{align}
Since $\min_{x} P(x) > \beta$ we have $P^L(x_1^L) > \beta^L$ for all $L$.  Let 
	\begin{align}
	\alpha = \frac{\mu_2(|\mc{X}|,L) \cdot \delta}{\beta^L}~,
	\end{align}
and set
	\begin{align}
	\hat{P} = (1 - \alpha) \tilde{P} + \alpha P^L~.
	\end{align}
Then $\hat{P}(x_1^L) > 0$ for all $x_1^L$ and by the triangle inequality:
	\begin{align}
	\maxvar{ \bar{P} }{ \hat{P} } 
	&\le \maxvar{ \bar{P} }{ \tilde{P} } + \maxvar{ \tilde{P} }{ \hat{P} } \\
	&< \mu_2(|\mc{X}|,L) \cdot \delta + \alpha \maxvar{ \tilde{P} }{ P^L } \\
	&< \left( 1 + \frac{1}{\beta^L} \right) \mu_2(|\mc{X}|,L) \cdot \delta~.
	\end{align}
Therefore for $\delta$ sufficiently small, we can choose a $\hat{P}$ such that $\maxvar{ \bar{P} }{ \hat{P} } < \epsilon$ for any $\epsilon > 0$.

\textbf{Case 2.}  We turn now to the second case.  Suppose that $\min_{x \in \mc{X}} P(x) = 0$.  Let $\mc{X}_0 = \{x \in \mc{X} : P(x) = 0
\}$ and $\mc{Z} = \mc{X} \setminus \mc{X}_0$.  Let $Q \in \mc{P}(\mc{Z})$ be the restriction of $P$ to $\mc{Z}$.  Then $Q$ is a probability distribution on $\mc{Z}$.  First suppose that $|\mc{Z}| = 1$.  Then $P(x) = 1$ for some $x \in \mc{X}$.  Let 
	\begin{align}
	\hat{P}(x_1^L) = P(x_1) \cdots P(x_L)~.
	\end{align}
Since all the marginal distributions $P_i$ of $\bar{P}$ satisfy $\maxvar{P}{P_i} < \delta$ we know that $\maxvar{\bar{P}}{\hat{P}} < \delta$.

Now suppose $|\mc{Z}| \ge 2$.  We can construct $\hat{P}$ by first finding a a joint distribution $\bar{Q}$ that is close to $\bar{P}$ and then invoking the first case of this proof on $\bar{Q}$.  From (\ref{eq:avg:closemarg}) we know that for some $c > 0$ we have
	\begin{align}
	\sum_{x_1^L \notin \mc{Z}^L} \bar{P}(x_1, x_2, \ldots, x_L) 
	&\stackrel{\Delta}{=} c \delta \\
	&< |\mc{X}|^L \delta~.
	\end{align} 
Define $\bar{Q}$ by
	\begin{align}
	\bar{Q}(x_1^L) = \left\{
		\begin{array}{ll}
		\bar{P}(x_1^L) + |\mc{Z}|^{-L} c\delta 
			& x_1^L \in \mc{Z}^L \\
		0 & x_1^L \notin \mc{Z}^L
		\end{array}
		\right.
	\end{align}
Since $\bar{Q}$ has support only on $\mc{Z}^L$ we can think of it either as a distribution on $\mc{X}^L$ or on $\mc{Z}^L$.  Note that
	\begin{align}
	\maxvar{ \bar{P} }{ \bar{Q} } < c \delta~.
	\end{align}
Let $\{Q_i : i \in [L]\}$ be the $i$-th marginal distributions of $\bar{Q}$:
	\begin{align}
	Q_i(x_i) = \sum_{x_j : j \ne i} \bar{Q}(x_1, x_2, \ldots, x_L) = Q_i(x_i) \qquad \forall i,\ x_i \in \mc{Z}~.
	\end{align}
Then we have for some $c' > 0$
	\begin{align}
	\maxvar{ Q }{ Q_i } < c' \delta~.
	\end{align}

Now we can apply Case 1 of this proof using the set $\mc{Z}$ and distributions $Q$, $\{Q_i\}$, and $\bar{Q}$.  For any $\epsilon_1 > 0$ we can find a $\delta_1 > 0$ such that if $\{ Q_i \}$ satisfy
	\begin{align}
	\maxvar{Q}{Q_i} < \delta_1~,
	\end{align}
then there exists a $\hat{Q}$ with marginals equal to $Q$ such that
	\begin{align}
	\maxvar{ \bar{Q} }{ \hat{Q} } < \epsilon_1~.
	\end{align}
Let $\hat{P}$ be the extension of $\hat{Q}$ to a distribution on $\mc{X}^L$ by setting $\hat{P}(x_1^L) = \hat{Q}(x_1^L)$ for $x_1^L \in \mc{Z}^L$ and $0$ elsewhere.  By the triangle inequality we have
	\begin{align}
		\maxvar{ \bar{P} }{ \hat{Q} }
		&\le \maxvar{ \bar{P} }{ \bar{Q} } + \maxvar{ \bar{Q} }{ \hat{Q} } \\
		&< c \delta + \epsilon_1~.
	\end{align}
We can choose $\delta$ sufficiently small so that $\delta_1$ and $\epsilon_1$ are sufficiently small to guarantee that this distance is less than $\epsilon$.
\end{proof}

\begin{lemma}
\label{lem:avg:symP}
Let $\mc{W}$ be an AVC with state cost function $\scostf(\cdot)$ and constraint $\scostb$ and let $L$ be a positive integer.  Let $\epsilon > 0$ be arbitrary and suppose $P$ is a distribution with $\symcost_L(P) < \scostb - \epsilon$.  Then there exists a $\delta > 0$ and $n_0$ such that for any $(n,N,L)$ list code with $n \ge n_0$  and $N \ge L+1$ whose codewords $\{\mbf{x}(i) : i \in [N] \}$ satisfy
	\begin{align}
	\maxvar{ \typ{\mbf{x}(i)} }{ P } &< \delta \qquad \forall i \in [N] \\
	\symcost_L( \typ{ \mbf{x}(i) } ) &< \scostb - \epsilon \qquad \forall i \in [N]~,
	\end{align}
the average error for the code is lower bounded: 	
	\begin{align}
	\max_{\mbf{s} \in \mc{S}^n(\scostb)} \listavgerr(\mbf{s}) > \frac{1}{L+1} - \frac{L}{N(L+1)}~.
	\end{align}
\end{lemma}

\begin{proof}
From Lemma \ref{lem:dist_approx} we can see that for any $\epsilon_1 > 0$ there exists a $\delta_1 > 0$ such that for any set $J \subset [N]$ of codewords with $|J| = L$ and $\maxvar{ \typ{\mbf{x}(j)} }{ P } < \delta_1$, we can find a joint type $\bar{P} \in \mc{P}(\mc{X}^L)$ with marginals equal to $P$ such that the joint type $\typ{ \mbf{x}(J) }$ satisfies
	\begin{align}
	\maxvar{ \typ{\mbf{x}(J)} }{ \bar{P} } < \epsilon_1~.
	\end{align}
Now let $U$ achieve the minimum in the definition of $\symcost_L(P)$.  Since $\symcost_L(P) < \scostb - \epsilon$ we have
	\begin{align}
	\sum_{s, x_1^L} \scostf(s) U(s | x_1^L) \typ{ \mbf{x}(J) }( x_1^L )
	&\le \sum_{s, x_1^L} \scostf(s) U(s | x_1^L) \bar{P}(x_1^L)
		+ \epsilon_1 \scostmax |\mc{X}|^{L} \\
	&< \scostb - \epsilon + \epsilon_1 \scostmax |\mc{X}|^{L}~,
	\end{align}
where $\scostmax = \max_{s \in \mc{S}} l(s)$.  Now choose $\epsilon_1 = \epsilon/(2 \scostmax |\mc{X}|^{L})$ so that
	\begin{align}
	\sum_{s, x_1^L} \scostf(s) U(s | x_1^L) \typ{ \mbf{x}(J) }( x_1^L )
	&< \scostb - \epsilon/2~,
	\end{align}
and choose $\delta = \delta_1$ according to Lemma \ref{lem:dist_approx}.

The jammer will pick a $J \subset [N]$ with $|J| = L$ uniformly from all such subsets and select its state sequence according to the random variable $\mbf{S}(J)$ with distribution
	\begin{align}
	Q^n(\mbf{s}) = \prod_{t=1}^{n} U( s_t | \{x_t(j) : j \in J\})~.
	\end{align}
The expected cost of $\mbf{S}(J)$ is
	\begin{align}
	\frac{1}{n} \expe[ \scostf(\mbf{S}(J)) ] 
		&= \frac{1}{n} \sum_{t=1}^{n} \sum_{\mbf{s}} 
			l(s_t) U( s_t | \{x_t(j) : j \in J\}) \\
		&= \sum_{s, \tilde{x}^L} 
			l(s) U( s | \tilde{x}_1, \ldots, \tilde{x}_L ) 
			\frac{ |\{t : x_t(j) = \tilde{x}_j\ \forall j\}| 
				}{
				n
				} \\
		&= \sum_{s, \tilde{x}^L} 
			l(s) U( s | \tilde{x}_1^L ) \typ{ \mbf{x}(J) } \\
		&< \scostb - \epsilon/2~.
	\end{align}
We can also bound the variance of $l(\mbf{S}(J))$:
	\begin{align}
	\mathrm{Var} \left( l(\mbf{S}(J)) \right) \le \frac{(\scostmax)^2}{n}~.
	\end{align}
Then Chebyshev's inequality gives the bound:
	\begin{align}
	\prob( \scostf( \mbf{S}(U_J,J) ) > \scostb ) 
		&\le 
		\frac{(\scostmax)^2}{n (\scostb - (\scostb - \epsilon/2) )^2} \\
		&\le
		\frac{4 (\scostmax)^2}{n \epsilon^2}~.
	\label{eq:avg:conv:cheby1}
	\end{align}

We now need some properties of symmetrizing channels used with the random variables $\mbf{S}(J)$.  Firstly, we have:
	\begin{align}
	\expe \left[ W^n( \mbf{y} | \mbf{x}(i), 
			\mbf{S}(J) ) \right] 
	&=
	\sum_{\mbf{s}} W^n( \mbf{y} | \mbf{x}(i), 
			\mbf{s}) U^n( \mbf{s} | \{x(j) : j \in J\}) \\
	&= 
	\expe \left[ W^n( \mbf{y} | \mbf{x}(j), 
			\mbf{S}(J \setminus \{j\} \cup \{i\} ) )
			\right]~.
	\label{eq:avg:conv:symswap}
	\end{align}
Using (\ref{eq:avg:conv:symswap}) we can see that for some subset $G \subset [N]$ with $|G| = L+1$:
	\begin{align}
	\sum_{i \in G}
		\expe \left[ 
			\listavgerr(i, \mbf{S}(G\setminus\{i\}))
			\right]
	&= \sum_{i \in G} 
		\left(1 - \sum_{\mbf{y} : i \in \ldec(\mbf{y})} 
		\expe \left[ 
			W^n(\mbf{y} | \mbf{x}_i, \mbf{S}(G\setminus\{i\}) )
		\right] \right)  \\
	&= L + 1 - \sum_{i \in G} \sum_{\mbf{y} : i \in \ldec(\mbf{y})}
		\expe \left[ W^n(\mbf{y} | \mbf{x}_{i_0}, \mbf{S}_{G \setminus \{i_0\}}) 		\right]~.
	\end{align}
Because each $\mbf{y}$ can be decoded to a list of size at most $L$	, we can get a lower bound
	\begin{align}
	\sum_{i \in G} \expe \left[ 
		\stdmaxerr(i, \mbf{S}_{G \setminus \{i\}}) \right] 
	&\ge L + 1 - L \sum_{\mbf{y} \in \mc{Y}^n} 
		\expe \left[ W^n(\mbf{y} | \mbf{x}_{i_0}, \mbf{S}_{G \setminus \{i_0\}}) 	\right] \nonumber \\
	&= 1~.
	\end{align}

We can now begin to bound the probability of error for this jamming strategy.   Let $\mc{J}$ be the set of all subsets of $[N]$ of size $L$, and let $\mbf{J}$ be a random variable uniformly distributed on $\mc{J}$.  We can write the expected error as
	\begin{align}
	\expe_{\mbf{J},\mbf{S}(\mbf{J})} \left[ 
		\listavgerr(\mbf{S}(\mbf{J})) \right]
	& 
	= \frac{1}{\binom{N}{L}} \frac{1}{N} 
		\sum_{J \in \mc{J}} \sum_{i=1}^{N}
		\expe \left[ \listavgerr(i, \mbf{S}(J)) \right]~.
	\end{align}
Then we have:
	\begin{align}
	\expe_{\mbf{J},\mbf{S}(U_{\mbf{J}},\mbf{J})} \left[ 
		\listavgerr(\mbf{S}(U_{\mbf{J}},\mbf{J})) \right]
	&
	\ge \frac{1}{\binom{N}{L}} \frac{1}{N}
		\sum_{G \subset [N]: |G| = L+1}
		\sum_{i \in G}
		\expe \left[ 
			\listavgerr(i, \mbf{S}(G\setminus\{i\})) 
			\right]~.
	\label{eq:avg:conv:firstlb}
	\end{align}
Now we can rewrite the inner sum using (\ref{eq:avg:conv:symswap}):
	\begin{align}
	\expe_{\mbf{J},\mbf{S}(\mbf{J})} \left[ 
		\listavgerr(\mbf{S}(\mbf{J})) \right] 
	&\ge \frac{ \binom{N}{L+1}
		}{ \binom{N}{L} \cdot N } \\
	&= \frac{ \binom{N}{L} \frac{N-L}{L+1} }{ \binom{N}{L} \cdot N }\\
	&= \frac{N - L}{(L+1) N} \\
	&= \frac{1}{L+1} - \frac{L}{N(L+1)}~.
	\end{align}
Finally, we can add in the bound (\ref{eq:avg:conv:cheby1}) to obtain
	\begin{align}
	\frac{1}{L+1} - \frac{L}{N(L+1)} 
	&\le 
		\expe_{\mbf{J},\mbf{S}(\mbf{J})} \left[ 
		\listavgerr(\mbf{S}(\mbf{J})) \right] \\
	&\le
	 \max_{\mbf{s} \in \mc{S}^n(\scostb)} \listavgerr(\mbf{s})
	 + \prob\left( l(\mbf{S}(\mbf{J})) > \scostb \right) \\
	&\le 
	 \max_{\mbf{s} \in \mc{S}^n(\scostb)} \listavgerr(\mbf{s}) \frac{4 (\scostmax)^2}{n \epsilon^2}~.
	\end{align}
Now, we can choose $n_0$ large enough such that 
	\begin{align}
	 \max_{\mbf{s} \in \mc{S}^n(\scostb)} \listavgerr(\mbf{s}) > \frac{1}{L+2} - \frac{L}{N(L+1)}~.
	\end{align}
\end{proof}

\begin{lemma}
\label{lem:avg:symAllP}
Let $\mc{W}$ be an AVC with state cost function $\scostf(\cdot)$ and constraint $\scostb$ and let $L$ be a positive integer.  For any $\epsilon > 0$ there exists a $\nu(L,\mc{W},\epsilon) > 0$ and $n_0$ such that for any $(n,N,L)$ list code $(\lenc,\ldec)$ with $n \ge n_0$  and $N > L+1$ whose codewords $\{\mbf{x}(i) : i \in [N] \}$ satisfy
	\begin{align}
	\symcost_L( \typ{ \mbf{x}(i) } ) &< \scostb - \epsilon \qquad \forall i \in [N]~,
	\label{eq:avg:symallP:cost}
	\end{align}
the error must satisfy
	\begin{align}
	\max_{\mbf{s} \in \mc{S}^n(\scostb)} \listavgerr(\mbf{s}) > \nu(L,\mc{W},\epsilon)~.
	\end{align}
\end{lemma}

\begin{proof}
Fix $\epsilon > 0$.  For each $P \in \mc{P}(\mc{X})$ from Lemma \ref{lem:dist_approx} we know there is a $\delta(P) > 0$ such that any joint distribution $\bar{P}$ with marginals within $\delta(P)$ of $P$ can be approximated by a $\hat{P}$ with marginals equal to $P$ such that $\maxvar{\bar{P}}{\hat{P}} < \epsilon$.  Let
	\begin{align}
	\mc{B}(P) = \left\{
		P' \in \mc{P}(\mc{X}) : \maxvar{P}{P'} < \delta(P)
		\right\}~.
	\end{align}
Then $\{\mc{B}(P) : P \in \mc{P}(\mc{X})\}$ is an open cover of $\mc{P}(\mc{X})$.  Since $\mc{P}(\mc{X})$ is compact there is a constant $r$ and finite subcover  $\{\mc{B}(P_j) : j \in [r]\}$.  From this finite cover we can create a partition $\{A_j : j \in [r]\}$ of $\mc{P}$ such that $A_j \subseteq \mc{B}(P_j)$ for all $j$.

Now consider an $(n,N,L)$ code whose codewords $\mc{C}$ satisfy (\ref{eq:avg:symallP:cost}).  Let $F_j = \{i \in [N] : \typ{\mbf{x}(i)} \in A_j \}$.  We can bound the error
	\begin{align}
	\listavgerr(\mbf{s}) 
	= \frac{1}{N r} \sum_{j = 1}^{r} \sum_{i \in F_j}
		\listavgerr(i,\mbf{s})
	\ge \frac{|F_j|}{N r} \left( \frac{1}{|F_j|} \sum_{i \in F_j}
		\listavgerr(i,\mbf{s}) \right)~.
	\end{align}
Since $\{F_j\}$ partition the codebook, for some $j$ we have $|F_j| \ge N/r$.  From Lemma \ref{lem:avg:symP} the jammer can force the error to be lower bounded by
	\begin{align}
	\max_{\mbf{s} \in \mc{S}^n(\scostb)} \listavgerr(\mbf{s}) 
		\ge 
		\frac{1}{r^2} \left( \frac{1}{L+1} - \frac{L}{N(L+1)} \right)~.
	\end{align} 
Since the constant $r$ is a function of $\epsilon$, $\mc{W}$ and $L$, we are done.
\end{proof}

Theorem \ref{thm:avg:converse} follows from the preceding Lemma.

\bibliographystyle{IEEEtranS}
\bibliography{listrefs}

\end{document}